\documentstyle[12pt,world_sci]{article}
\input epsf
\newcommand{\lsim}{\stackrel{<}{\sim}}
\begin{document}
\begin{flushright}
BNL-60809
\end{flushright}

\title{{\bf SUMMARY OF HEAVY ION THEORY}}
\author{SEAN GAVIN\thanks{This manuscript has been authored under contract
number DE-AC02-76CH00016 with the U.S. Department of Energy.}\\
{\em Physics Department, Brookhaven National Laboratory\\
Upton, NY, 11973 USA}}

\maketitle
\setlength{\baselineskip}{2.6ex}

\begin{center}
\parbox{13.0cm}
{\begin{center}
ABSTRACT
\end{center} {\small \hspace*{0.3cm}
Can we study hot QCD using nuclear collisions?  Can we learn about
metallic hydrogen from the impact of comet Shoemaker-Levy 9 on
Jupiter?  The answer to both questions may surprise you!  I summarize
progress in relativistic heavy ion theory reported at DPF `94 in the
parallel sessions.}}
\end{center}

%\section{Introduction}

Lattice simulations of QCD demonstrate that matter at temperatures
exceeding $T_c\sim$~150~MeV is very different from matter composed of
hadrons \cite{1}.  Simulations commonly display dramatic changes in
thermodynamic quantities, such as the energy density, in a narrow
interval $|T-T_c|\lsim 5$~MeV, indiciting an abrupt transformation
from hadronic to quark-gluon degrees of freedom.  The underlying aim
of the theoretical speakers in the heavy ion sessions has been to
understand how properties of high temperature matter can be deduced
from collisions of nuclei at RHIC and LHC at $\sqrt{s}=$~200 and
5500~GeV per nucleon, respectively.  Is the high temperature state
deconfined?  Is chiral symmetry restored?  Is the expected abrupt
transformation a true phase transition?  Physics demands {\em
experimental} answers to these questions.

Talks in this session addressed the global dynamics of heavy ion
collisions as well as specific probes of the high temperature state.
The significant progress in understanding the collision dynamics at
the Brookhaven AGS, $\sqrt{s}\sim 5$~$A$GeV, and the CERN SPS,
$\sqrt{s}\sim 20$~$A$GeV was surveyed by Schlagel and Vogt.  Sarcevic
and Shuryak discussed two important probes of the dynamics that will
be more important at higher energies: open charm and direct photon
production.  The suppression of $J/\psi$ production in ion-ion
collisions probes the deconfinement of the high temperature state.
This topic was presented by Satz and Thews.  Ayala and Petrides discussed
the modification of parton distributions in nuclei, a related topic.
Sch{\"a}fer and Shuryak discussed the nature of the chiral transition.
Disoriented Chiral Condensates, a possible probe of the dynamics of
chiral symmetry breaking at RHIC and LHC, was discussed by Kluger.

What can we learn about high temperature QCD from nuclear collisions
\cite{2}?  An analogous, similarly-complex question is, what can the
impact of a comet with Jupiter teach us about the equation of state of
hydrogen?  We have all seen exciting images of the collisions of the
fragments of the comet Shoemaker-Levy 9 with Jupiter that took place
from 16 to 22 July, 1994.  Observations of the impacts are providing
new information on comet structure and the stratification and
composition of Jupiter's atmosphere \cite{3}.  Indeed, the speed of
sound in metallic hydrogen \cite{4} can be measured if reflections of
downward-launched accoustic waves from Jupiter's core can be observed
\cite{5}.

\section{Collision Dynamics}

A very important issue in the AGS and SPS fixed-target experiments has
been the {\em stopping power}.  The extent to which the projectile ion
is stopped (`slowed' is more accurate) as it crashes through the
target nucleus depends on how the constituents interact.  Should one
treat the constituents as nucleons or quarks on the time scales of the
collision?  Is resonance formation important in the nucleon
rescattering?  Is there a formation time for secondary particle
production?

Data on stopping comes primarily from the rapidity distribution of
protons.  Schlagel showed that the AGS proton data for projectiles as
large as $Au$ can be described by a purely hadronic rescattering model
that incorporates resonance formation.  He also showed that the
omission of resonance formation does not describe the data.  We expect
formation time effects to become more important at the higher SPS
energy.  Vogt showed that SPS data for light projectiles can be
described by string models that incorporate these effects.  She argued
that the Pb beam runs commencing this fall will be useful in
deciding between string and hydrodynamic models.

In the case of Shoemaker-Levy 9, one is also interested in how the comet
is stopped by Jupiter's atmosphere.  The comet's structure determines
the depth to which the comet penetrates.  Stopping therefore provides
information on comet structure.  High temperature H$_2$O emission
lines that are likely from the comet remnants have been observed
\cite{6}.  In the following table, I list aspects of the dynamics in nuclear
collisions discussed at this meeting together with their analogs in
the comet-Jupiter impact.

\begin{center}
\begin{tabular}{|l||c|c|}        \hline
               &Ion + Ion                     &Comet + Jupiter \\ \hline\hline
stopping       &baryon distribution           &comet remnants   \\
               &proton $dN/dy$                &H$_2$O \\ \hline
thermalization &$\gamma, e^+e^-, \mu^+\mu^-$; &emission lines   \\
               &$\pi,\, K,\ldots, D$          &H$_2$S, CH$_4$   \\ \hline
flow           &directed flow\cite{7}         &plume, ejecta    \\ \hline
density        &$J/\psi, \psi', \Upsilon$     &absorption lines \\ \hline
EOS            &DCC                           &seismic waves    \\ \hline
\end{tabular}
\end{center}

Another central question in AGS/SPS nuclear collisions is {\em
thermalization}: how effectively is the momentum of the projectile
distributed among the participant nucleons and produced hadrons?  Do
these hadrons reach local thermal equilibrium and undergo collective
flow?  The abundance of produced particles such as pions, kaons and
antiprotons indicates the extent to which particles interact and
thermalize.  Shuryak observed that photon and dilepton production can
be used to measure the temperatures that the system achieves, although
backgrounds can be formidable in practice.  In the comet-Jupiter
collision, the excitation of high temperature emission lines provide
information on energy deposition.  The observation of hot H$_2$S
indicates that the comet penetrated below Jupiter's ammonia-rich cloud
cover into a layer containing ammonium hydrosulfide \cite{3}.

Collective flow is another proposed indirect indication of
thermalization.  The directed flow of primary and produced particles
in nuclear collisions can be deduced from exclusive measurements
\cite{7}.  The flow shown in Fig.~1 depends on particle type.  For
example, antiproton flow is anticorrelated with the proton flow
because ${\overline p}$'s can annihilate with $p$'s.  Observe that
directed flow at the high AGS/SPS energies has little to do with the
equation of state, EOS, since the eikonal approximation holds.  Flow
about the impact site of fragment G on Jupiter is shown in Fig. 2.
The ejecta are asymmetrically distributed along the direction the
impact and rich in CH$_4$, indicating that the matter is
from Jupiter.  Note that in both cases the flow comes about because
the collisions are not central.
\begin{figure}[t]
\epsfxsize=2.2in
\vspace{-30pt}
\centerline{\epsffile{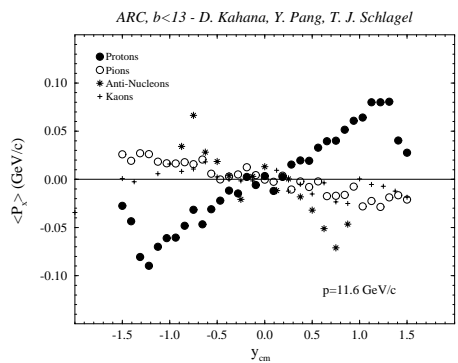}}
\vspace{-30pt}
\caption{
Directed flow for various particle species in Au+Au at the AGS from Ref. [7]}
\end{figure}
\begin{figure}[t]
\epsfxsize=3.0in
\centerline{\epsffile{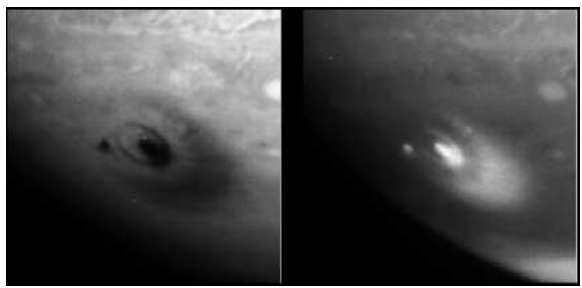}}
%\vspace{-36pt}
\caption{
Directed flow on Jupiter due to the impact of comet fragment G from
the Hubble Space Telescope, Ref [8]. Left and right images are taken
with green and methane filters respectively.}
\end{figure}

New dynamical questions emerge at the higher RHIC and LHC energies.
Minijets with transverse momenta larger than $p_{{}_T}\sim$ 2~GeV can
produce perhaps up to 50\% of particle production in Au+Au at RHIC.
The low-x rise in the parton distributions measured at HERA implies
that minijets can dominate particle production at LHC \cite{9}.
Minijets can produce a high density of partons at short times $\sim
{p_{{}_T}}^{-1}\sim 0.1$~fm.  These partons can rescatter and
thermalize forming what Shuryak calls `hot glue.'

Open charm production can serve as a measure of the temperature of
this hot glue.\cite{10} The point is that charm is heavy and hard to make
unless temperatures are very high.  Most of the charm production is
therefore expected to occur via primary hard perturbative scattering.
However, semihard $gg\rightarrow c\bar{c}$ rescattering in a hot
glue system can enhance charm production relative to perturbative
expectations.  Sarcevic pointed out that reliable perturbative
estimates are needed to provide a benchmark and presented
calculations to $O(\alpha_s^3)$.

\section{$J/\psi$ Suppression}

$J/\psi$ production provides a probe of the densities achieved in
nuclear collisions that is also sensitive to deconfinement.  Matsui
and Satz observed that a $J/\psi$ can exist in a low density quark
gluon plasma as a QCD Bohr atom.  However, color screening inhibits
the binding of the $c\bar{c}$ pair when the temperature $T$ is high enough
that the screening length $\propto T^{-1}$ is smaller than the $c\bar{c}$'s
Bohr radius.  The $c$ and $\bar{c}$ can then wander apart to
form open charm, leading to a suppression of the $J/\psi\rightarrow
\mu^+\mu^-$ peak relative to the dimuon continuum.

A cellestial analogy to $J/\psi$ suppression is the modification of
the intensity of hydrogen absorption lines in stars.  The degree of
ionization of hydrogen in the solar plasma depends on its temperature,
{\it i.e.} a hotter star has fewer H atoms and more ions than a cool
one.  Correspondingly, the intensities of H absorption lines are
reduced relative to the continuum in hot stars.  Many processes
contribute to the line spectra, so that detailed models are needed
extract the temperature from the line intensity \cite{11}.
Nevertheless, line intensities are now a well established method for
measuring stellar temperatures.

$J/\psi$ suppression holds similar promise as a density probe in
nuclear collisions, although its analysis is clearly much more
complicated.  While the production of the $c\bar{c}$ pair is
perturbative and calculable, the formation of the bound state is not.
Correspondingly, $J/\psi$ production is not well understood even in
$p\bar{p}$ collisions at the Tevatron (although there has been recent
progress at high $p_{{}_T}$ \cite{12}).  Thews presented a quantum
mechanical analysis of the spacetime evolution of the $c\bar{c}
\rightarrow J/\psi$.  Such an analysis is necessary for understanding
the formation of bound states in the high density environment.

In addition to the Matsui-Satz effect, there are several
``background'' contributions to $J/\psi$ suppression.  Although they
are interesting manifestations of QCD, these contributions make the
interpretation of density signals ambiguous.  Initial state parton
scattering broadens the $p_{{}_T}$ distributions in $pA\rightarrow
J/\psi + X$ and Drell Yan, and is more-or-less understood \cite{13}.
Ayala, Petridis and Sarcevic discussed the modification of parton
distributions in nuclei compared to free nucleons.  The resulting EMC
and parton-shadowing effects alter $J/\psi$ production, as Petridis
emphasized.  Final state scattering adds to the suppression effect, as
hadronic reactions like $N+J/\psi\rightarrow D\bar{D}N$ and $\rho+
J/\psi\rightarrow D\bar{D}$ can take place; see the presentations by
Satz and Vogt.

A new direction taken at this meeting has been to seek {\em
first-principles} constraints on models of the background
contributions.  Ayala reported on work with McLerran, Venugopalan and
Jalilian-Marian on the development of new theoretical tools for
calculating parton distributions in large nuclei.  Their idea is that
at small $x$, the QCD scale is determined by the number of partons per
unit transverse area, which varies as $A^{1/3}$.  For very large
nuclei, they formulate a weak-coupling semiclassical method to
calculate the parton distributions.  When perfected, these methods can
provide important constraints on models of parton shadowing.

Satz presented work with Kharzeev in which they argue that the total
$J/\psi$-nucleon cross section can be calculated.  Following Peskin
and Bhanot, they treated the heavy $c\bar{c}$ system as nearly
pointlike and applied a short-distance operator-product-expansion
analysis.  In principle, cross sections calculated by this method can
be used to constrain models of final state interactions.  Of course,
the fact that roughly half of hadroproduced $J/\psi$ come from
electromagnetic decays of relatively large $\chi$ states implies that
not all of the final state interactions are calculable.  Satz argued
that one can subtract the uncalculable $\chi \rightarrow J/\psi+
\gamma$ contribution by detecting the photon.

\section{Disoriented Chiral Condensate?}

Equilibrium high temperature QCD manifests a chiral symmetry if the
light up and down quarks are taken to be massless.  However, a phase
transition occurs at a critical temperature $T_c\sim 140$~MeV at
which chiral symmetry is broken by the formation of a scalar $\langle
{\overline q} q\rangle$ condensate.

Rajagopal and Wilczek\cite{14} pointed out that the chiral condensate
can be temporarily disoriented in the nonequilibrium environment of a
heavy ion collision.  Near $T_c$, the approximate chiral symmetry
implies that the scalar condensate is nearly equivalent to a
pion--like pseudoscalar isovector condensate $\sim
\langle {\overline q}\gamma_5{\vec\tau} q\rangle$, where $\vec\tau$
are the Pauli isospin matrices.  Consequently, domains containing a
macroscopic pion field can appear as the temperature drops below
$T_c$.  Such domains will eventually disappear as the system evolves
towards the true vacuum in which only the scalar condensate is
nonzero.

Bjorken, Kowalski, Taylor and others pointed out that DCCs can lead to
fluctuations in the charged and neutral pion spectra.\cite{15} In the
heavy ion system, the evolving DCC domains can radiate pions
preferentially according to their isospin content.  However, the
ability of experimenters to identify DCCs amidst the background
produced by conventional mechanisms critically
depends on the domains' size and energy content.\cite{16} At this
meeting, Kluger discussed efforts to calculate DCC formation using the
linear sigma model.  In this model, the the pion field is coupled to a
scalar $\sigma$ field that characterizes the scalar condensate $^1$.
The fields interact through the potential $V =
\lambda({\vec\pi}^2+\sigma^2-v^2)^2/4 - H\sigma$ that is intended
to describe the behavior of QCD near $T_c$.

Many agree \cite{16,17,18} that the scale of the domain size
is fixed by the inverse sigma mass $m_\sigma^{-1}\sim \{\lambda
v^2\}^{-1/2}$.  The question is, what is the value of $m_\sigma$ in
the high density system?  Kluger, Cooper, Mottola and Paz studied the
time evolution of the linear sigma model in a self consistent large
$N$ approximation, where $N$ is the number of pions.  Domains are
small in this model, perhaps $\sim 1-3$~fm, because $m_\sigma$ is
large at $T_c$.  Alternatively, M\"uller and I observed that if
$m_\sigma(T_c) = 0$ (as would be the case if chiral restoration were
strictly second order), domains would be much larger and, perhaps,
observable \cite{18}.

If seen in ion-ion collisions, DCC's can provide information about
the equation of state of hot QCD.  Similarly, the detection of seismic
waves on Jupiter may teach us about the EOS of metallic hydrogen.
Both effects are fascinating but may prove very difficult to observe!

But what about QCD?  The nature of the phase transition is unknown for
realistic values of the $u$, $d$ and $s$ quark masses.  The real
transition is likely continuous, but with the large increase in the
energy density mentioned earlier.  The linear sigma model does not
describe this increase.  Nevertheless, QCD can exhibit large
fluctuations in the transition region indicative of nearly critical
behavior as described by the three-flavor sigma model.\cite{19}
Sch\"afer and Shuryak suggest that an instanton liquid model may
capture both of these features.  Sch\"afer observed that the pion
suffers strong interactions at high $T$ as in the sigma
model.\cite{20} Shuryak argued that the instanton liquid model can
also explain the large energy density change in QCD.  Models like this
may therefore provide a more realistic context for studying dynamical
phenomena such as DCC's than the linear sigma model.

To summarize, there has been substantial progress in understanding the
hard-core phenomenology of Au+Au at the AGS and S+Au at the SPS.  The
heavy ion experimental program is driving towards heavier projectiles
and higher energies, with Pb+Pb at the SPS this fall and RHIC at
$\sqrt{s} = 200$~$A$GeV in 1999.  Fascinating phenomena are expected
and their complicated backgrounds are coming to be understood.
There is every reason to keep looking up!

I thank L. Bildsten, J. Milana, B. M\"uller, R. D. Pisarski, A. Stange,
and F. Weber for helpful discussions.

\bibliographystyle{unsrt}

\end{document}